\documentclass[a4paper]{article}

\usepackage{INTERSPEECH2016}

\usepackage{graphicx}
\usepackage{amssymb,amsmath,bm}
\usepackage{textcomp}

\usepackage{hyperref}
\usepackage{multirow}
\usepackage{comment}

\DeclareMathOperator{\len}{len}
\DeclareMathOperator{\KLD}{KLD}

\ninept

\title{Automatic Genre and Show Identification of Broadcast Media}

\makeatletter
\def\name#1{\gdef\@name{#1\\}}
\makeatother \name{{\em  Mortaza Doulaty, Oscar Saz, Raymond W. M. Ng, Thomas Hain}}

\address{Speech and Hearing Group (SpandH), Department of Computer Science, University of Sheffield \\
  {\small \tt \{mortaza.doulaty, o.saztorralba, wm.ng, t.hain\}@sheffield.ac.uk }
}

\begin{document}
	\maketitle

 	\begin{abstract}
% TH Morrie that was almost a rewrite,
% you really should look at the style on how this is done normally ... 
Huge amounts of digital videos are being produced and broadcast every day, leading to giant media archives. 
Effective techniques are needed to make such data accessible further. Automatic meta-data labelling of 
broadcast media is an essential task for multimedia indexing, where it is standard to use multi-modal input 
for such purposes. This paper describes a novel method for automatic detection of media genre and show 
identities using acoustic features, textual features or a combination thereof.
Furthermore the inclusion of available meta-data, such as time of broadcast, is shown to lead to very high performance. 
Latent Dirichlet Allocation is used to model both acoustics and text, yielding fixed dimensional representations of 
media recordings that can then be used in Support Vector Machines based classification. Experiments are 
conducted on more than 1200 hours of TV broadcasts from the British Broadcasting Corporation (BBC), where the 
task is to categorise the broadcasts into 8 genres or 133 show identities. On a 200-hour test set, accuracies of 
98.6\% and 85.7\% were achieved for genre and show identification respectively, using a combination of 
acoustic and textual features with meta-data.
	\end{abstract}

	\section{Introduction}
	\label{sec:introduction}
  	% Introduction
With the ever increasing amounts of digital media and requirements
to process media archives, automatic labelling and classification
of media recordings becomes more and more important. Multimedia data can
be grouped by genre such as sports, news and comedy, which are
categories that also imply other than purely semantic information. 
As such classification is easier to understand by viewers, is required for 
%With these high-level information, 
%multimedia data can be organised in a much more efficient way in 
downstream processes such as indexing. Research in this field is pushed forward 
by initiatives such as the ``MediaEval Benchmarking for Multimedia Evaluation" 
\cite{MediaEval}, or the ``Robust, as Accurate as Human Genre Classification for Video" 
challenges within the Multimedia Grand Challenges of ACM Multimedia Conference \cite{mmgc2009}. 
Genre identification, and identification of shows can be considered as a core task in multimedia processing and is studied in this paper.

%With the ever increasing amounts of digital media and requirements to process media archives, automatic labelling and classification of them becomes more and more important. In multimedia indexing, genre identification is an essential task where it is standard to use multi-modal input for such purposes and mostly audio and video are being used \cite{montagnuolo2007tv}. Other types of data such as text and meta-data can also be used to further improve the classification performance \cite{ekenel2013multimodal}. Research in this field is further being motivated by initiatives such as ``MediaEval Benchmarking for Multimedia Evaluation" \cite{MediaEval} and ``Robust, as Accurate as Human Genre Classification for Video" challenges within the Multimedia Grand Challenges of ACM Multimedia Conference \cite{mmgc2009}. %\footnote{http://comminfo.rutgers.edu/conferences/mmchallenge/}

In a typical genre ID setting supervised methods learn from audio and/or video features extracted from the media streams.
%In a typical genre ID task, audio and video features from the multimedia are extracted on which supervised learning of genre is performed. 
For audio-based classification mostly short-term features are used \cite{liu98}, such as Mel-Frequency Cepstral
Coefficients (MFCC) \cite{roach2001classification}. The use of other features such as average
speech rate, signal energy, zero crossing rate, duration of silence, noise and speech have also been studied \cite{ekenel2013multimodal}. Typical features extracted from video include colour statistics, camera motion and cut detection \cite{ekenel2013multimodal,montagnuolo2009parallel,mironica2013depth}.
% Moved from below to here
In the literature, audio based features usually have very
similar performance compared to the
video-based features \cite{kim2013line}.
% check whether the below is valid: MD: Yes
Textual features such as subtitles and meta-data (e.g. title, tag, video description) contain
semantic information and are believed to give promising
results in genre ID \cite{ekenel2013multimodal}. 
%Nevertheless, to the best of our knowledge they are not fully exploited.

%For audio based features, mostly mel frequency cepestral coefficients are used. Some other features such as average speech rate, duration of silence, noise and speech can also be used \cite{ekenel2013multimodal}. In case of video, colour stats, camera motion, cut detection and etc. are used \cite{ekenel2013multimodal,montagnuolo2009parallel}.

%In the literature, these features are usually grouped into structural and cognitive features. Most of the features named so far are considered as structural features. Average number of faces in the screen in video based features can be considered as a cognitive feature. 
%In this work audio based features are mostly studied and compared with textual features. Furthermore, the use of meta-data is also studied.

This paper proposes new methods for automatic detection of media genre based on audio and explores what information sources are required to obtain very high levels of performance on a very large dataset of more than 1,200 hours of data. Also for the first time, to the best of our knowledge, the show identification task on very large datasets is studied in this paper.

This paper is organised as follows: Section 2 reviews the related work for genre identification. Section 3 describes the proposed method for genre and show identification, followed by the experimental setup in Section 4, results in Section 5 and a conclusion of this work in Section 6.
    
	\section{Related Work}
	\label{sec:relatedwork}
  	Research on genre ID tasks typically report accuracies of over 90\% \cite{ekenel2013multimodal,montagnuolo2009parallel,kim2013line,montagnuolo2007tv}. Typical datasets are the RAI dataset \cite{montagnuolo2007tv}, Quaero dataset \cite{quaero} and some custom YouTube videos. Both RAI and Quaero datasets are around 70 hours each and most of other datasets have similar or smaller sizes. 

Genre labelling is difficult even for humans, mostly because of its subjectiveness. 
%Labels of genres in the datasets sometime are not very coherent and this adds to the complexity of this task. 
Labels of genres differ among datasets and this makes interpretations of results difficult. Also, the chosen labels do not always fully reflect multi-genre TV; for instance the RAI dataset has 7 genres labels. These 7 genres are cartoon, commercial, football,
music show, news, talk show and weather forecast, which seem
to be in some cases very specific, e.g. football which can be
considered as a subset of a broader sport genre. 

The proposed method in \cite{kim2013line} uses acoustic
features and using the RAI dataset, they
reported accuracy of 94.3\%. Using video, 99.2\% was reported
in \cite{ekenel2013multimodal} for the same dataset. For other similar datasets such as the Quaero dataset, similar classification
accuracies are reported (e.g. 94.5\% \cite{ekenel2013multimodal}). 
% I would not say 89.7% is qualitatively different from 95%.
% Because we don't want to let reviews think what we did
% is qualitatively different (worse) than those 90% accuracy work
On a custom YouTube dataset \cite{ekenel2013multimodal}, 87.3\% was reported which was further improved by the use of meta-data to 89.7\%

%Typically accuracies of over 90\% are reported for genre ID tasks \cite{kim2013line,ekenel2013multimodal,montagnuolo2007tv,montagnuolo2009parallel}. The proposed method in \cite{kim2013line} uses acoustic features and using the RAI dataset \cite{montagnuolo2007tv} with 70h of data, they reported accuracy of 94.3\%. Using video, 99.2\% was reported in \cite{ekenel2013multimodal} for the same dataset. Other similar datasets with the same size are also used, such as Quaero dataset with similar classification accuracies being reported (e.g. 94.5\% \cite{ekenel2013multimodal}). Lower accuracies are also reported on a custom YouTube dataset \cite{ekenel2013multimodal} which was further improved by the use of meta-data from 87.3\% to 89.7\%.

Genre ID can be addressed by using
generative models. Kim et al. \cite{kim2013line} reported an accuracy of 93.6\% on a 11.5h test set with the RAI dataset using Gaussian
Mixture Models (GMM) trained with the MFCC features. These features represent
short-term characteristics of speech, such as the spectral properties
of phonemes and speakers. In smaller and more homogeneous
datasets where the same shows and speakers might often
reoccur, the classification performance with those features
are usually much better than the accuracies obtained on larger and more heterogeneous datasets \cite{saz2014slt}. 

The probabilistic approach using GMMs can be further extended
using latent semantic indexing techniques. 
\cite{kim2013line} had the accuracy improved by 0.7\% absolute over their GMM baseline of 93.6\% on the RAI dataset
using acoustic topic models. They used vector quantisation
to represent frames by discrete symbols and trained Latent
Dirichlet Allocation (LDA) models \cite{kim2009acoustic} followed by Support
Vector Machine (SVM) classifiers.
%into 7 genres. These 7 genres are cartoon, commercial, football,
%music show, news, talk show and weather forecast, which seem
%to be in some cases very specific, e.g. football which can be
%considered as a subset of a broader sport genre. 
However when the amount of data is more and thus the dataset is more diverse, the same baseline models performs much worse \cite{saz2014slt}.

%Kim et al. \cite{kim2013line} reported a baseline accuracy of 93.6\% using Gaussian Mixture Models (GMM) trained with Mel-Frequency Cepstral Coefficient features on the RAI dataset. These short-term spectral features represent short-term characteristics of speech, like the spectral properties of phonemes and speakers. In smaller and more homogeneous datasets where the same shows and speakers might often reoccur, the classification performance with these features are usually much better than larger and more heterogeneous datasets \cite{saz2014slt}. They further improved the accuracy by 0.7\% absolute using acoustic topic models where they used vector quantisation to represent frames by discrete symbols and trained Latent Dirichlet Allocation (LDA)  models followed by Support Vector Machine (SVM) classifiers to classify the RAI dataset into 7 genres. These 7 genres are cartoon, commercial, football, music show, news, talk show and weather forecast, which seem to be in some cases very specific, e.g. football which can be considered as a subset of a broader genre. However when the amount of data is more and thus the dataset is more diverse, the same baseline models performs much worse \cite{saz2014slt}.

Sageder et al. \cite{sageder2016group} tried to pool various types of features and then group and select a subset using canonical correlation analysis in order to identify low-correlated and complementary features. These features were then used to train different classifiers such as K-Nearest Neighbour, Random Forest and SVM. They reported very good classification performance on different datasets including some custom RAI and BBC shows, however the amount of data is tiny (less than 55h in total and in case of BBC, 4.5h with just 3 classes) and thus hard to directly compare with other approaches.

Other approaches try to identify certain audio-visual events, with the objective to model the semantics of the broadcast shows or YouTube videos \cite{lee2010audio, castan2013indexing}. However, due to the complexity of the shows and videos, the performance of these techniques are not usually competitive with the previously mentioned methods.

%The work in this paper aims to provide a novel approach for automatic detection of media genre and show identities using acoustic features, textual features and a combination of both. Furthermore the inclusion of meta-data such as time of broadcast and channel information is studied. The amount of data used in these experiments are unlike other papers where usually less than few hundred hours of data were used. Here more than 1200 hours of data is used and it imposed some unique challenges which are discussed later. 
%Also in terms of number of classes most other papers had around 7 classes, where in an extra task here show 

\begin{comment}
tv does not come with much meta data
\end{comment}

    \section{Acoustic Latent Dirichlet Allocation}
    \label{sec:acclda}
	% Acoustic LDA
As shown in our previous work \cite{doulaty2015ldadnn}, acoustic LDA domain posteriors have a unique distribution across genres and shows. In this work we make use of acoustic LDA domain posterior features to classify broadcast media and investigate the use of other data sources such as subtitles, automatic speech recognition (ASR) output as well as meta-data. 

LDA is an unsupervised probabilistic generative model for collections of discrete data. Since speech observations are continuous data, first it needs to be represented by some discrete symbols, here called acoustic words. A GMM with $N$ mixture components is employed for this purpose. The index of Gaussian component with the highest posterior probability is then used to represent each frame with a discrete symbol. Frames of every acoustic document of length $T$, $ \mathbf{d}_i = \{\mathbf{u}_1,...,\mathbf{u}_t,...,\mathbf{u}_T\} $ are represented as:
\begin{equation}
v_t=\underset{n}{\arg\max} \; P(G_n | \mathbf{u}_t), \;\; 1 \le n \le N
\end{equation}
Where $G_n$ is a Gaussian component from a mixture of $N$ components. With this new representation, document $\mathbf{d}_i$ is represented as $\tilde{\mathbf{d}}_i = \{v_1,...,v_t,...,v_T\} $.
For each acoustic word $v_t$ in each acoustic document $\tilde{\mathbf{d}}_i$, %raw-counts are computed and based on their document frequencies, 
term frequency-inverse document frequency (tf-idf) can be computed as:
\begin{equation}
w_t = tfidf(v_t, \tilde{\mathbf{d}}_i, \tilde{\mathbf{D}}) = tf(u_t, \tilde{\mathbf{d}}_i) \; idf(u_t, \tilde{\mathbf{D}})
\end{equation}
Where $\tilde{\mathbf{D}}$ is the set of all acoustic documents represented with acoustic words. With each document now represented with tf-idf scores as $\bar{\mathbf{d}}_i = \{w_1,...,w_t,...,w_T\}$, the LDA models can be trained.

\begin{figure}
	\centering
	\includegraphics[width=6cm]{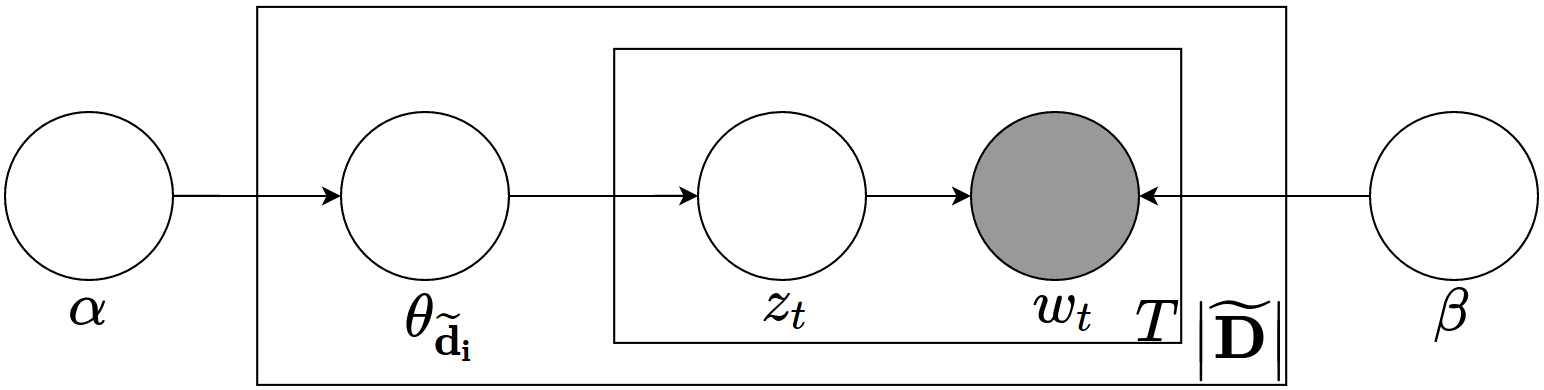}
	\caption{Graphical model representation of LDA}
	\label{fig:lda-graphical-model}
\end{figure}

A graphical representation of the LDA model is shown at Figure \ref{fig:lda-graphical-model}, as a 
three-level hierarchical Bayesian model. In this model, the only observed variables are $w_t$'s. $\alpha$ and $\beta$ are dataset level parameters, $\theta_{\mathbf{\tilde{d}_i}}$ is a document level 
variable %and $z_t$, $\bar{x}_t$ are symbol level variables. 
and $z_t$ is a latent variable indicating the domain from which $w_t$ was drawn. The following joint distribution is the result of the generative process of LDA:
\vspace{-1mm}
\begin{equation}
p(\theta, \mathbf{z}, \mathbf{\bar{d}} | \alpha, \beta) 
= p(\theta | \alpha) \prod_{t=1}^{T}p(z_t | \theta) p(w_t|z_t,\beta)
%= p(\theta | \alpha) \; p(\mathbf{z} | \theta) \; p(\mathbf{\bar{d}}|\mathbf{z},\beta)
\end{equation}
The posterior distribution of the latent variables given the acoustic document and $\alpha$ and $\beta$ parameters is:
\vspace{-1mm}
\begin{equation}
\label{eq:posterir}
p(\theta, \mathbf{z} | \mathbf{\bar{d}}, \alpha, \beta) = 
\frac{p(\theta, \mathbf{z}, \mathbf{\bar{d}} | \alpha, \beta)}{p(\mathbf{\bar{d}} | \alpha, \beta)}
\end{equation}
Computing $p(\mathbf{\bar{d}} | \alpha, \beta)$ requires some intractable integrals. A reasonable approximate can be acquired using variational approximation, which is shown to work reasonably well in various applications \cite{blei2003latent}. The approximated posterior distribution is:
\vspace{-1mm}
\begin{equation}
\label{eq:approx-posterior}
q(\theta, \mathbf{z} | \gamma, \phi) = q (\theta | \gamma) \prod_{t=1}^{T}q(z_t | \phi_t)
\end{equation}
where $\gamma$ is the Dirichlet parameter that determines $\theta$ and $\phi$ is the parameter for the multinomial that generates the latent variables. 
%Figure \ref{fig:lda-approx-graphical-model} shows the graphical representation of the approximated topic model.

Training minimises the Kullback-Leiber Divergence between the real and the approximated joint probabilities (equations \ref{eq:posterir} and \ref{eq:approx-posterior}) \cite{blei2003latent}: 
\vspace{-1mm}
\begin{equation}
\underset{\gamma, \phi}{\arg\min} 
\; \KLD \big(
q(\theta, \mathbf{z} | \gamma, \phi)
\; || \; 
p(\theta, \mathbf{z} | \mathbf{\bar{d}}, \alpha, \beta)
\big)
\end{equation}
\vspace{-1mm}

The posterior Dirichlet parameter $\gamma(\mathbf{\bar{d}})$ can be used as feature for classification. Discriminative classifiers such as SVMs have been used successfully for genre classification tasks before \cite{kim2013line, rouvier2009factor} including our previous work \cite{saz2014slt}.

Kim et al. \cite{kim2013line} used the whole shows to train the LDA models and used the domain posteriors as features for an SVM classifier. In this work we followed our previous setup \cite{doulaty2015ldadnn,doulaty2015lda} where only speech segments are used to train the LDA model. For each show, the domain posteriors of its segments were accumulated and length normalised and used as features for the discriminative classifier in the later stage:

\begin{equation}
\mathbf{x}_i = \frac{1}{\sum\limits_{s \in segs} \len(s)} \sum_{i \in segs} \len(i) \; \gamma(\mathbf{\bar{d}}_i)
% or \displaystyle\sum_{}
\end{equation}

	\section{Experimental Setup}
	\label{sec:exp}
	% Experimental Setup
\subsection{Data}

\begin{figure}[t]
	\centering
	\includegraphics[width=8cm]{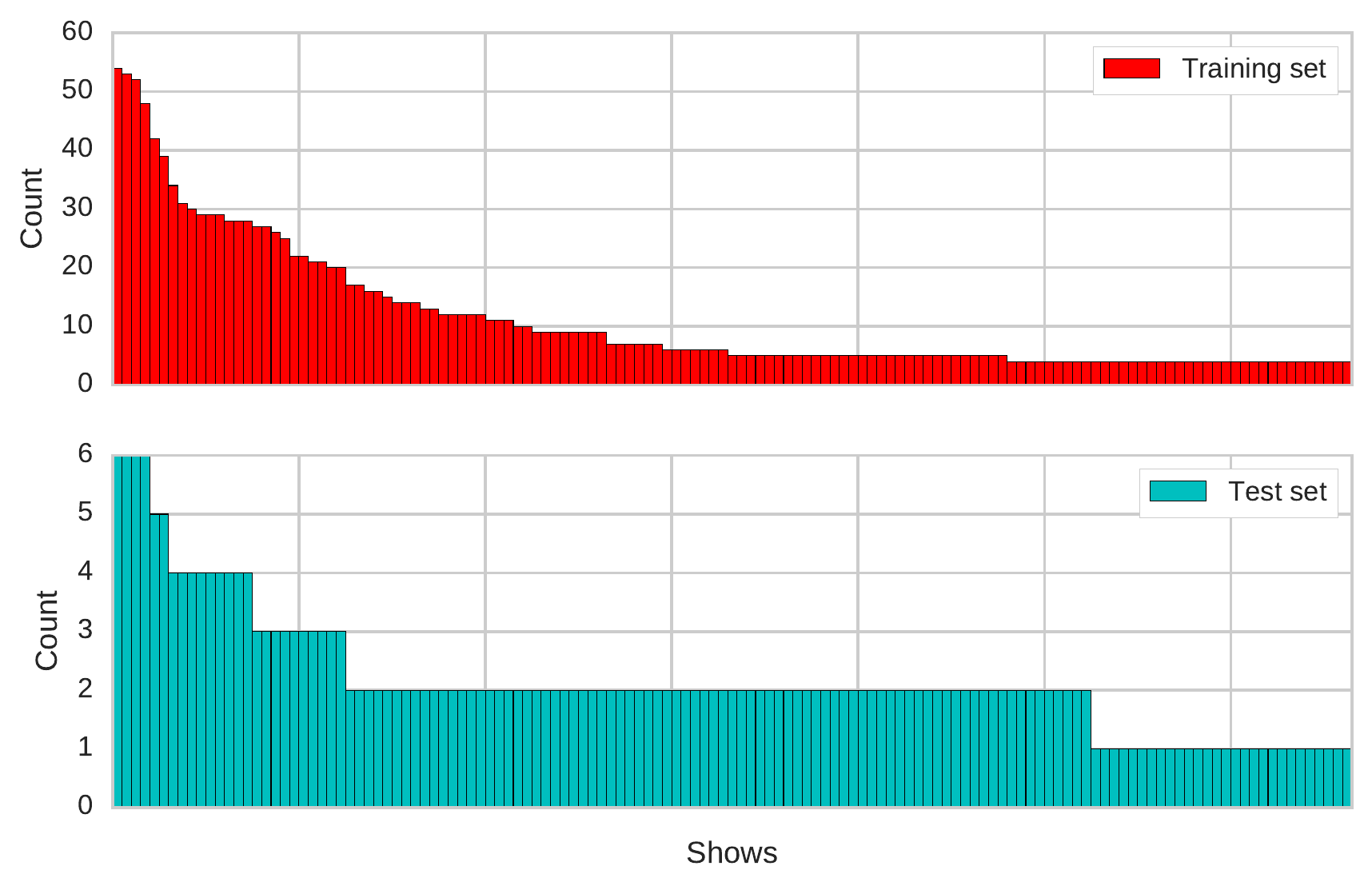}
	\vspace{-8mm}
	\caption{Distribution of 133 unique shows in training and test set}
	\label{fig:show-dist}
\end{figure}

TV broadcasts provided by the British Broadcasting Corporation (BBC) were used for all experiments. The data is identical to the one defined and provided for the 2015 Multi-Genre Broadcast (MGB) Challenge \cite{MGB} with a different training/testing set definitions. The shows were chosen to cover the full range of broadcast show types and categorised in 8 genres: advice, children's, comedy, competition, documentary, drama, events and news. All shows were broadcast by the BBC during 6 weeks in April and May 2008. There were more then 2,000 shows in the original MGB challenge data, from which 1,789 shows were selected for the experiments, 1,501 shows for the training set and 288 shows for test set, with 133 unique shows in total. The distribution of shows (time and count) across genres for the training and test data is shown in Table \ref{tab:data}. Figure \ref{fig:show-dist} shows the distribution of the 133 unique shows for both training set and test set, where the horizontal axis represents unique shows and the vertical axis represents the number of episodes in that show. Order of the bars are identical in both plots and e.g. the first bar of both plots represents the same show.

%where vertical axis is the counts of episodes and horizontal axis is the unique shows. Bars in each plots correspond a unique show   On average there are 11 episodes of the same show in the training set and 2 episodes in the test set. 
It is important to note that this dataset is by orders of magnitude larger than most of the datasets used in the literature for the genre ID task \cite{mmgc2009,roach2001classification,ekenel2013multimodal,montagnuolo2009parallel,kim2013line,sageder2016group}.
% roach2001classification can be deleted

\begin{table}
\centering
\caption{Amount of training and testing data per genre}
\label{tab:data}
\begin{tabular}{|l|c|c|c|c|}
\hline
\multicolumn{1}{|c|}{\multirow{2}{*}{Genres}} & \multicolumn{2}{c|}{Train Set}  & \multicolumn{2}{c|}{Test Set} \\ \cline{2-5} 
\multicolumn{1}{|c|}{}                        & \# Shows      & Dur (h)         & \# Shows     & Dur (h)        \\ \hline \hline
Advice                                        & 189           & 135.3           & 35           & 24.4           \\ \hline
Children's                                    & 301           & 112.7           & 60           & 25.0           \\ \hline
Comedy                                        & 90            & 44.1            & 22           & 10.8           \\ \hline
Competition                                   & 224           & 153.3           & 45           & 29.8           \\ \hline
Documentary                                   & 90            & 57.4            & 29           & 19.3           \\ \hline
Drama                                         & 102           & 69.0            & 21           & 14.6           \\ \hline
Events                                        & 98            & 161.0           & 21           & 36.3           \\ \hline
News                                          & 407           & 293.0           & 55           & 40.2           \\ \hline
\textbf{Total}                                & \textbf{1501} & \textbf{1025.6} & \textbf{288} & \textbf{200.4} \\ \hline
\end{tabular}
\end{table}

\begin{comment}
\begin{figure}[h]
	\centering
	\includegraphics[width=8cm]{images/show-dist.pdf}
	\caption{Distribution of 133 unique shows in training and testing set}
	%\label{fig:show-dist123}
\end{figure}
\end{comment}

% % % % % % % % % % % % % % % % % % % % % % % % % % % % % % % % %
\subsection{Baseline}
As a baseline, GMM classifiers were used for both genre and show identification tasks. For the data as described above, genre ID task has 8 target classes and show ID task has 133 target classes. 13 dimensional PLP \cite{hermansky1990perceptual} features plus their first and second derivatives were used to train the genre-based and show-based GMMs using Expectation Maximisation algorithm and mix-up procedure to reach 512 mixtures. The optimal number of mixtures for a similar task was found to be 512 in our previous experiments \cite{saz2014slt}. Table \ref{tab:gmmbaseline} shows the classification accuracy for both tasks. Since there are fewer target classes, genre ID should be an easier classification task compared to show ID. However, GMMs are found to perform better in classifying shows than genres ($70.1\%$ compared to $61.5\%$), one reason for this could be the diversity of data as discussed in the introduction and the fact that PLP features are good for representing speaker specific characteristics \cite{saz2014slt} and for the show ID task the GMMs are learning speakers in re-occurring episodes. However they provide poor generalisation for the genre ID task. If show to genre mapping is assumed to be \textit{a priori} knowledge, then the show ID GMMs can be used for the genre ID task. The accuracy for genre ID in such a setting would be $79.2\%$.

\begin{table}
\centering
\caption{Genre/show classification accuracy with GMMs}
\label{tab:gmmbaseline}
\begin{tabular}{|c|c|c|}
\hline
Model & Genre ID    & Show ID \\ \hline
GMM   & 61.5 (79.2) & 70.1    \\ \hline
\end{tabular}
\end{table}

\section{Results}
\subsection{Whole Show and Segment Based Acoustic LDA}
Whole shows were used to train the LDA models with varying number of latent domains with the same procedure outlined in the previous section. The performance of these models is to be compared with the proposed segment based LDA models. The classification accuracy for the genre ID and show ID tasks are presented in Table \ref{tab:acousticlda}. 
For the segment level models the posterior estimates on short segments can be noisy. Picking the domain with the highest posterior probability and representing the posterior vector as one-hot-vector may reduce the posterior estimate noise and it was found to slightly outperform the base case and was used in the experiments.
%Please note that for the segment level models, domain posteriors are converted to one-hot-vectors based on the posterior with the highest value which was found to slightly outperform the case where posteriors values are not filtered. 

As the performance of segment level models was better than the whole show models, they were used in the rest of experiments. Segment based models also had higher accuracy with fewer latent domains. % compared to the whole show models. 
E.g. the highest accuracy with the segment based models for genre ID was 86.4\% obtained with an LDA model with 256 latent domains. However, the best performance for the whole show models was 80.6\%, with 2048 latent domains. A similar pattern was found for the show ID task as well.

\begin{table}
\centering
\caption{Genre/show classification accuracy using whole show and segment based acoustic LDA models}
\label{tab:acousticlda}
\begin{tabular}{|c|c|c|c|c|}
\hline
\multirow{2}{*}{\#Domains} & \multicolumn{2}{c|}{Whole Show} & \multicolumn{2}{c|}{Segment Based} \\ \cline{2-5} 
                          & Genre ID        & Show ID       & Genre ID         & Show ID         \\ \hline \hline
16                        & 73.6            & 45.1          & 76.7             & 46.7            \\ \hline
32                        & 71.9            & 53.8          & 81.5             & 57.8            \\ \hline
64                        & 78.1            & 56.6          & 81.2             & 63.4            \\ \hline
128                       & 77.8            & 56.9          & 83.3             & 66.6            \\ \hline
256                       & 76.4            & 58.0          & 86.4             & 67.3            \\ \hline
512                       & 80.2            & 61.8          & 85.0             & 66.7            \\ \hline
1024                      & 77.1            & 65.3          & 85.7             & 63.8            \\ \hline
2048                      & 80.6            & 65.3          & 84.7             & 63.1            \\ \hline

\end{tabular}
\end{table}

\subsection{Text Based LDA}
%There should be some information in the text level
Transcripts of the shows have valuable information for discrimination of genres and shows. In this section the classification is studied based on solely textual features. BBC TVs provide subtitles of the TV soundtrack, mostly for helping deaf and hard-of-hearing viewers. The quality of these subtitles varies considerably by genres. For example subtitles of live events and news are mostly re-spoken live ASR output and have higher errors, however for other genres which does not have the live nature, the quality is higher. For a detailed analysis of the subtitles quality refer to  \cite{MGB} and \cite{MGB_ASR_SHEF}. Subtitles were used as-is, without any preprocessing, to train the classifiers for both tasks. 
Although subtitles can be of varied quality, their correctness is still high. In a second experiment, ASR output is used instead of subtitles.
%Furthermore to simulate the case where there is no sub-title available and to study the effect of ASR errors on the accuracy of classification tasks, ASR output text is used in an other experiment. 
The ASR systems used here were trained for participation in the MGB Challenge. For more details about these ASR systems, refer to \cite{MGB_ASR_SHEF} and \cite{MGB_DIA_SHEF}.
%the models trained for 
%Here we studied the provided subtitles as is (without any processing) and also used one of our ASR models trained for the MGB Challenge \cite{MGB_ASR_SHEF,MGB_DIA_SHEF} to transcribe the shows to simulate the scenario where no subtitle is available. 
The classification task here is similar to a document classification task, where each show's transcript is a document and the classes are either genres or shows. To have a fair comparison with the acoustic LDA experiments, text based LDA models were trained and the domain posteriors were used as features in the SVM classifiers. A simpler approach would be SVMs with tf-idf features directly. However here the LDA model reduces the dimensionality of the tf-idf features to the number of latent domains, which is known to work better than tf-idf only features for document classification \cite{blei2003latent}. Table \ref{tab:textlda} summarises the results. LDA models trained with the subtitles performed substantially better than models trained on the ASR output. Note that the ASR models used here have around 30\% WER on the official development set of the MGB challenge. The performance gap is even wider in case of the show ID task, 22.6\% vs. 13.5\% absolute difference. This could caused by some specific names that were present in the subtitles, but not in the ASR output. Such words may have considerable discriminability information.

The overall performance of text based classification with subtitles is generally better than with direct audio based classification (96.2\% vs. 84.4\% for the genre ID task and 81.3\% vs. 67.3\% for the show ID task) but when considering the ASR output only, the audio based classification is better for the show ID task. %but for the genre ID, text based is still more accurate.

\begin{table}
\centering
\caption{Genre/show classification accuracy using text based LDA models}
\label{tab:textlda}
\begin{tabular}{|c|c|c|c|c|}
\hline
\multirow{2}{*}{\#Domains} & \multicolumn{2}{c|}{Subtitles} & \multicolumn{2}{c|}{ASR Output} \\ \cline{2-5} 
                          & Genre ID       & Show ID       & Genre ID        & Show ID       \\ \hline \hline
16                        & 77.4           & 41.3          & 70.1            & 29.2          \\ \hline
32                        & 81.3           & 50.7          & 71.9            & 34.0          \\ \hline
64                        & 85.4           & 62.1          & 81.6            & 45.8          \\ \hline
128                       & 89.2           & 68.8          & 87.5            & 55.2          \\ \hline
256                       & 91.0           & 77.1          & 88.2            & 65.6          \\ \hline
512                       & 91.0           & 76.7          & 87.9            & 63.9          \\ \hline
1024                      & 94.8           & 81.3          & 88.5            & 64.9          \\ \hline
2048                      & 96.2           & 79.9          & 89.9            & 64.9          \\ \hline
4096                      & 93.1           & 78.1          & 89.6            & 64.2          \\ \hline
\end{tabular}
\end{table}

\subsection{Using Meta-Data}
The data used in the experiments also includes some meta-data, such as the BBC broadcast channel number, the date and time of broadcast, and other unstructured information. Using some of the structured meta-data is studied next to learn how the classification accuracy can be improved further. Since these programmes were broadcast during 6 weeks in April and May 2008, using the date was not likely to be helpful which we verified in the experiments. Instead, the time of broadcast, splitting 24 hours into 8 chunks, and channel number, in this setup 1--4 corresponding to BBC1, BBC2, BBC3 and BBC4, were appended as one-hot-vectors to the inputs of the SVM classifiers and their effect is studied. Table \ref{tab:metadata} summarises the results of using the meta-data together with acoustic LDA features. Adding these meta-data helps for both tasks. When comparing channel and time, in both tasks appending time helps more and the difference is bigger in case of the show ID task (72.8\% vs. 77.7\%). Combining channel information and time of broadcast also helps further improve the classification accuracy in both tasks and overall with meta-data there is 5.9\% and 15.3\% absolute improvement in accuracies of genre ID and show ID tasks. The first row in Table \ref{tab:metadata} shows the accuracy when only meta-data is used (without any acoustic or textual features) which shows how much information with the meta-data is provided.
%The same experiments repeated with textual features and since the improvements were not noticeable, they are not reported here. 

\begin{table}
\centering
\caption{Genre/show classification accuracy using meta-data}
\label{tab:metadata}
\begin{tabular}{|l|c|c|}
\hline
\multicolumn{1}{|c|}{Meta-Data} & Genre ID & Show ID \\ \hline \hline
Only Channel \& Time            & 46.7     & 22.0    \\ \hline
Baseline (acoustic 256)         & 86.4     & 67.3    \\ \hline
+ Channel                       & 89.6     & 72.8    \\ \hline
+ Time                          & 89.9     & 77.7    \\ \hline
+ Channel \& Time               & 92.3     & 82.6    \\ \hline
\end{tabular}
\end{table}

\subsection{System Fusion}
With the two systems based on acoustic and textual features, one can use a combination of both, assuming that they will make different classification errors and their outputs are complimentary. To combine the scores of the systems, logistic regression is used to find a linear combination of individual system scores to maximize the probability of correct classification \cite{FoCal}. Table \ref{tab:fusion} shows the classification accuracy with the system fusion. The combination of acoustic and text based systems improves the classification accuracy for both tasks, 97.2\% and 85.0\% accuracy for genre ID and show ID respectively, which shows the complementarity of the individual systems. Moreover, including meta-data further improves the accuracy to 98.6\% and 85.7\% which is near perfect for the genre ID task.

\begin{table}
	\centering
	\caption{Genre/show classification accuracy with system fusion}
	\label{tab:fusion}
	\begin{tabular}{|l|c|c|}
		\hline
		\multicolumn{1}{|c|}{Method}   & Genre ID & Show ID \\ \hline \hline
		Baseline (acoustic 256)            & 86.4     & 67.3    \\ \hline
		Baseline (text 2048)                 & 96.2     & 79.9    \\ \hline
		Acoustic \& Text                       & 97.2     & 85.0   \\ \hline
		Acoustic + Meta-data \& Text & 98.6     & 85.7    \\ \hline
	\end{tabular}
\end{table}

	\section{Conclusions}
	\label{sec:conclusion}
	In this paper new methods for the genre classification of broadcast media based on audio were proposed. Furthermore, required information sources to obtain very high levels of performance was explored. Also for the first time, show classification task on very large datasets was studied.
%In this paper a novel approach for genre and show identification of broadcast media is proposed which uses acoustic, textual and meta-data to reach very high level of  classification performance. 
For the experiments more than 1,200 hours of data with more than 1,500 TV shows from the BBC which was broadcast in 2008 was used. These data was a part of the MGB 2015 challenge \cite{MGB}. For the genre ID task there were 8 classes and for the show ID task there were 133 classes. Acoustic and textual LDA models were trained with the audio and subtitles to infer the posterior Dirichlet parameters which were then used in SVM classifiers to classify the genres and shows. On a 200h test set, combination of both acoustic and text based classifiers had accuracy of 97.2\% and 85.0\% for genre ID and show ID tasks respectively. Use of meta-data such as time of broadcast further improved the accuracies to 98.6\% and 85.7\%.

Future work can be exploiting more information from the unstructured meta-data and trying to deal with cases where some meta-data is missing.
	
	\section{Acknowledgements}
	This work was supported by the EPSRC Programme Grant EP/I031022/1 (Natural Speech Technology). The audio and subtitle data used for these experiments was distributed as part of the MGB Challenge 
	(\url{mgb-challenge.org}) \cite{MGB} through a licence with the BBC. 
	%\newpage
	\eightpt
	\bibliographystyle{IEEEtran}

	\bibliography{refs}

\end{document}